\documentclass[12pt]{article}

\usepackage{epsfig}

\newcommand{\la}[1]{\label{#1}}
\newcommand{\re}[1]{\ (\ref{#1})}
\newcommand{\nn}{\nonumber}
\newcommand{\ed}{\end{document}}
\newcommand{\be}{\begin{equation}}
\newcommand{\ee}{\end{equation}}
\newcommand{\ba}{\begin{eqnarray}}
\newcommand{\ea}{\end{eqnarray}}
\newcommand{\baz}{\begin{eqnarray*}}
\newcommand{\eaz}{\end{eqnarray*}}
\newcommand{\bb}{\begin{thebibliography}}
\newcommand{\eb}{\end{thebibliography}}
\newcommand{\ct}[1]{${\cite{#1}}$}

\begin{document}
\sloppy
\thispagestyle{empty}
\hfill {FTUV 99/32 ; IFIC 99/34}

\vspace{2cm}

\mbox{}

\begin{center}
{\large\bf   Instantons and  $\eta$ 
Meson Production  near threshold in NN collisions}

\vskip 0.5cm

 N.I. Kochelev$^{1}$ , V.Vento$^2$, A.V. Vinnikov$^{3}$\\
\vspace{5mm}
{\small\it
(1) Bogoliubov Laboratory of Theoretical Physics,\\
Joint Institute for Nuclear Research,\\
Dubna, Moscow region, 141980 Russia\\
(2) Departament de F\'{\i}sica Te\`orica and Institut de F\'{\i}sica 
Corpuscular,\\
Universitat de Val\`encia-CSIC\\ 
E-46100 Burjassot (Valencia), Spain \\
(3) Far Estern State University, Department of Physics,\\
Sukhanova 8, GSP, Vladivostok, 690660  Russia}

\vspace*{1cm}

\end{center}
\begin{abstract}
An enhancement for the $\eta$ production in
proton-neutron collisions as compared with that in
proton-proton scattering has been recently observed.
We present a calculation for the production cross section,
in proton-neutron collisions  near threshold, within instanton model
for the QCD vacuum and show that a specific flavor dependent 
nonperturbative  quark-gluon interaction related to instantons  
is able to explain the observed enhancement. 
\end{abstract}

\vskip 1.0cm
\leftline{Pacs: 12.39-x, 13.60.Hb, 14.65-q, 14.70Dj}
\leftline{Keywords:  neutron, proton, meson, quark, gluon, instanton, 
threshold.}
\vspace{0.5cm}

{\tt
\leftline {kochelev@thsun1.jinr.ru}
\leftline{vicente.vento@uv.es}
\leftline{vinnikov@thsun1.jinr.ru}
}

\newpage

\section{Introduction}

The interest of the investigation of meson production in nucleon-nucleon 
and nucleon-antinucleon collisions near threshold  has increased recently 
since different
anomalies  have been detected \ct{anom},\ct{eta},\ct{nbarn},\ct{cel}.
These  anomalies are characterized by unexpected enhancements 
in the production of the mesons which contain large admixtures of 
strange quarks in their wave function. The large  
OZI rule violations near threshold challenge  most of 
the theoretical approaches to the dynamics of strangeness production
in strong interactions (see discussion in \ct{ellis}).

One of this anomalous behaviors is related to the surprisingly large  cross 
section  in proton-neutron 
collisions near threshold  for the $\eta$ production \ct{eta},\ct{cel}. 
It has been found recently \ct{cel} that the ratio to the same 
process in p-p scattering in the same kinematical regime is about $7$. 

An  explanation of this enhancement,  based on 
the large contribution from $N^*(1535)$ ($S_{11}$) resonance has been
 suggested \ct{theory}. However the final result is very sensitive to the 
values of the meson-$S_{11}$ coupling constants, which at present are
not  well known. It is therefore difficult to establish
the precise contribution from the $S_{11}$ to this cross section and
confirm the validity of the mechanism.

In here, we present a different explanation for the enhancement, which
in our case is associated with a particular microscopic mechanism of
QCD. In the vacuum of QCD there exist strong fluctuations of the gauge
fields called instantons  (see recent review \ct{shur}). 
The instantons induce flavor dependent
quark-quark \ct{thooft} and quark-gluon interactions \ct{koch1}
whose structure is related with  the flavor and helicity 
properties of the quark zero-modes in the presence of the instanton
fields.
The most famous of them is t'Hooft's quark-quark interaction \ct{thooft} which 
has been successfully applied to describe the properties of the
QCD vacuum and hadron spectroscopy \ct{shur}.
Besides this interaction, which contains $2N_f$ quark legs, 
the instantons can also lead  to  different
type of the vertices which include arbitrary numbers of quark and 
gluon legs. The gluons  in these vertices produce easily
strange quarks. Therefore these new type of interactions should lead 
to large violations of the OZI rule in the strong interaction.
One  example of them, which  leads to a 
non-zero value of the chromomagnetic moment of the quark, has been 
discussed recently \ct{koch1}.

In Section 2 we discuss the effective quark-multi-gluon Lagrangian induced 
by the instantons. In Section 3  the specific contribution of the
mechanism  to the cross section for the  production of $\eta$
mesons in proton-neutron collisions is estimated. Finally in  
Section 4 the main conclusions of the present work will be drawn.

\section{ Effective quark-gluon vertices induced by instantons}   

The effective Lagrangian induced by instantons has the following
form~\cite{review}
\footnote{ For anti-instantons one should interchange $R\leftrightarrow L$.}
 \begin{eqnarray}
{\cal L}_{inst}&=&\int\prod_q(m_q\rho-2\pi^2\rho^3\bar q_R(1+\frac{i}{4}
\tau^aU_{aa^\prime}\bar\eta_{a^\prime\mu\nu}\sigma_{\mu\nu})q_L)
\nonumber\\
&\cdot &exp({-\frac{2\pi^2}{g}\rho^2U_{bb^{\prime}}
\bar\eta_{b^\prime\gamma\delta}
G^b_{\gamma\delta}})
\frac{d\rho}{\rho^5}d_0(\rho)d\hat{o},
\label{e1}
\end{eqnarray}
 where $\rho$ is the instanton size, $\tau^a$ are the Pauli
matrices associated with the generators of the
$SU(2)_c$ subgroup of the $SU(3)_c$ color group,
 $d_0(\rho)$ is the density of the instantons, $d\hat{o}$ stands
 for integration over the instanton orientation in color space,
$\int d\hat{o}=1$,
$U$ is the orientation matrix of the instanton,
 $\bar\eta_{a\mu\nu}$ represent t'Hooft's symbols. This Lagrangian
describes the effective interaction between gluons and quarks arising
from QCD through the color dipole moment of the instanton.

The effective quark--gluon interaction which generates the
contribution  to $\eta$ production can be obtained by expanding
(\ref{e1}) in  powers of the gluon strength and integrating over
$d\hat{o}$. The effective parameter of this expansion is the ratio 
of the average instanton size $\rho_c$ to the confinement radius
$R_{conf}$.   This ratio, in realistic models for the QCD vacuum
\ct{shur}, is rather  small, $\rho_c^2/R_{conf}^2\approx 0.1$,  and
therefore we can restrict the expansion  to the lowest order. The
corresponding approximation contains contributions to the production
of mesons with  definite quantum numbers.

 To second order in $G_{\gamma\delta}^b$ we found out that
effective Lagrangian, which is responsible for $\eta$
production, has the following form 
\begin{equation}
\Delta L_{\eta}=-\int{\cal L}_{t'Hooft}
\frac{\pi^3\rho^4d\rho}{8\alpha_s(\rho)} G_{\mu\nu}^a \widetilde{G}_{\mu\nu}^a,
\label{e2}
\end{equation}
where ${\cal L}_{t'Hooft}$ is  t'Hooft's interaction which for 
massless $u$ and $d$ quarks  is

${\cal L}_{t'Hooft} = $
\be
\frac{16\pi^4d_0(\rho)\rho}{9}
(\bar u_{R}u_{L}\bar d_{R}d_{L}+\frac{3}{8}(
\bar u_{R}t^au_{L}\bar d_{R}t^ad_{L}-\frac{3}{4}\bar
u_R\sigma_{\mu\nu}t^a
\bar u_L\bar d_R\sigma_{\mu\nu}t^ad_L)),
\label{e44}
\ee
This quark-gluon vertex is presented in Fig.1.
\begin{figure}[htb]
\centering
\epsfig{file=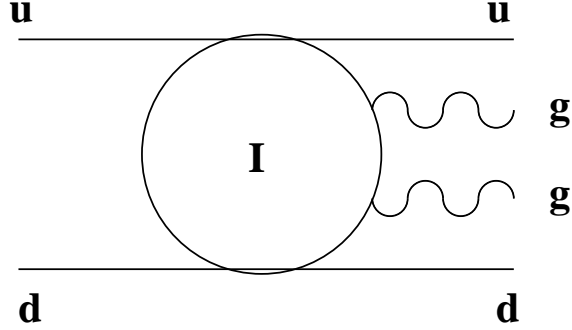,width=7.5cm}
\vskip 0.5cm
\caption{\it The effective quark-gluon vertex which gives rise to
the contribution in $\eta$ meson production.}
\end{figure}
Within the simplest version of instanton liquid model for the QCD vacuum, 
the size of instantons is fixed
\begin{equation}
d_0(\rho)\sim\delta(\rho-\rho_c).
\label{e3}
\end{equation}
$\rho_c\sim 1.6\div 2.0 \; GeV^{-1} $ is the average size of the 
instantons  in the QCD vacuum. This value is obtained from hadron 
spectroscopy \ct{shur}. However this size 
is  not very well defined, for example, recent lattice calculations 
\ct{lattice} give a value for it which is larger, 
$\rho_c\sim 2.5\; GeV^{-1}$.
We will therefore in our calculation  consider a large interval for
$\rho_c\sim 1.6\div 2.5 \; GeV^{-1} $.

By using the low energy theorem, based on the chiral anomaly, one can
calculate the following matrix element \ct{voloshin}, \ct{shifman} 

\be
<0|\frac{3\alpha_s}{4\pi}G_{\mu\nu}^a \widetilde{G}_{\mu\nu}^a|\eta>\approx
\sqrt{3}f_\pi m_\eta^2,
\la{low}
\ee
where we have neglected $m_u$ and $m_d$ with  respect to $m_s$.
Under the assumptions of ref. \ct{shifman} 
the quark-$\eta$ meson interaction from \re{e2} becomes
\be
 {\cal L}_{\eta}= \frac{2\pi^6\rho_c^6f_\pi m_\eta^2}{\sqrt{3}
\alpha_s^2(\rho_c)}
 (\bar u_{R}u_{L}\bar d_{R}d_{L}+\frac{3}{8}(
\bar u_{R}t^au_{L}\bar d_{R}t^ad_{L}-\frac{3}{4}\bar
u_R\sigma_{\mu\nu}t^a
\bar u_L\bar d_R\sigma_{\mu\nu}t^ad_L)) \phi_\eta.
\la{et1}
\ee
where $\phi_\eta$ is the $\eta$-field.
It should be mentioned that this effective interaction  can only be used 
if the frequency of gluon fields is  small. This condition however
holds for the production of the particles near threshold.

\section{ Contribution of the instantons to the $\eta$ production
cross section near threshold}

A very unique feature of the interaction \re{et1} is its 
strong flavor dependence. Indeed it is non zero only if quarks have different 
flavors. Therefore, one can expect that the contribution of  
it to the $\eta$ meson production in proton-neutron 
collision is larger then in proton-proton and 
neutron-neutron ones, because in the former  the flavor structure
favors the probability of interaction.

The contribution of the instanton interaction is determined by 
the matrix element of interaction \re{et1} 
\be
<NN|{\cal L}_\eta|NN\eta>\propto<NN|{\cal L}_{t'Hooft}|NN>.
\la{mat} 
\ee
Due to the strong spin-flavor dependence of t'Hooft's interaction 
\re{e44}, the final result is very sensitive to the spin-flavor
structure of the nucleon wave function. We will use a model for hadron
structure which has proved to be consistent with the instanton picture
and which will favor the proposed mechanism.

It is well known 
that a significant violation of $SU(6)$ symmetry 
in the nucleon wave function is required to explain some properties of 
nucleons: $\Delta$-nucleon mass splitting, enhancement of $u$ 
quark distribution with respect to 
$d$ quark distribution at large Bjorken $x$ region in deep-inelastic
scattering, behavior of nucleon form factors, etc... (see \ct{diquark}).
It was shown \ct{koch2} that the instanton induced 
interaction can explain the nucleon-$\Delta$ mass difference, 
due to the  formation of a quasi-bound scalar $ud$
diquark inside nucleon. At present, the instanton induced 
diquark configuration is widely under discussion, not only 
in  connection with hadron spectroscopy, but also  
as a possible scenario for the formation of a color superconducting
state  in quark-gluon matter \ct{wilczek}.
Therefore we feel encouraged to use this simplifying scenario.

In the quark-diquark model, proton and neutron consist of a scalar $ud$ diquark 
and a u-quark or d-quark, respectively.
It is evident that  due to specific flavor properties of the 
instanton vertex only the
interaction between quarks with different flavors, which are not 
included into   diquark configuration,  can lead to the $\eta$ meson
production in nucleon-nucleon scattering (see Fig.2)
\footnote{The diquark-quark interaction mediated by one instanton 
vanishes, as determined by
 t'Hooft's mechanism, since the two quarks that participate in the
interaction have the same flavor.
A diquark-quark interaction, through this mechanism, arises only
if mediated by at least two {\it different} instantons, and 
is suppresed by the smallness of the density of instantons in the QCD vacuum.}.
\begin{figure}[htb]
\centering
\epsfig{file=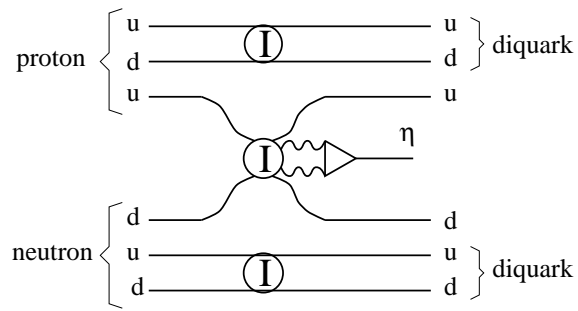,width=7.5cm}
\vskip 0.5cm
\caption{\it The instanton contribution to the  $\eta$
meson  production in proton-neutron interaction}
\end{figure}
Therefore the instanton induced $\eta$ meson  production in proton-proton 
interaction within  pure 
quark-$ud$-diquark model should be very small in comparison with 
proton-neutron scattering \footnote{ If we neglect the production of
the $\eta$ meson in the diquark-quark interaction and contributions 
from other mechanisms ($S_{11}$, ...) then the ratio of cross sections
becomes $\sigma_{pp}/\sigma_{pn}\approx 0$.}.  The experimental observation of a 
sizeable 
enhancement of the $\eta$ production in proton-neutron scattering,        
gives arguments in the favor of a large  diquark component in the 
nucleon wave function, however the proposed dynamical mechanism favors
the enhancement in other scenarios, e.g., bag model or naive quark
model, as well \footnote{In the case of the pure 
$SU(6)$ nucleon wave function one can expect the ratio
$\sigma_{pp}/\sigma_{pn}\approx 1/4$, based on the result of  
\ct{farrar} where the single instanton contribution to the ratio of 
amplitudes of elastic pp and pn scattering has been obtained,
$\Phi(pp)/\Phi(pn)=-8/17$.}. 

Let us to estimate the cross section for $\eta$-meson production 
in proton-neutron collisions. We follow the notation:
\be 
p(p_1)+n(p_2)\rightarrow p(p_1^\prime)+n(p_2^\prime)+\eta(p_\eta).
\label{reaction}
\ee
The cross section is given by
\be
d\sigma=\frac{dPS^3}{4\sqrt{(p_1.p_2)^2-m_p^4}}
\sum_{spin}|M|^2,
\la{cross}
\ee
where
\be
dPS^3=\frac{d^3p_1^\prime d^3p_2^\prime d^3p_\eta}{(2\pi)^92E_1^\prime
2E_2^\prime 2E_\eta}(2\pi)^4
\delta(p_1+p_2-p_1^\prime-p_2^\prime-p_\eta),
\la{ps}
\ee
is the 3-body phase space volume.
$M$ is the matrix element  of the interaction \re{mat} and
$\sum_{spin}$ represents the spin summation over the final and
spin averaging over the initial nucleon states.

Only the colorless part of the instanton induced interaction can 
contribute to the cross section of the reaction \re{reaction}.
Within the quark-diquark model one can factorize the matrix element
and obtains using a nonrelativistic approximation for 
the quark wave functions of the nucleon,

$M\sim <PN|\bar u_Ru_L\bar d_Rd_L|PN>\sim <P|\bar u_Ru_L|P><N|\bar
d_Rd_L|N>$
\be
\approx\bar P_RP_L\bar N_RN_L.
\label{mat2}
\ee

The final result for the matrix element is
\footnote{We took into account in \re{cross2} the additional 
factor two related to the contribution of anti-instantons.}
\be
\sum_{spin}|M|^2=\frac{\pi^{12}\rho_c^{12}f_\pi^2m_\eta^4}{54\alpha_s^4
(\rho_c)}
(8m_p^4-2m_p^2(t_1+t_2)+t_1t_2)F_N(t_1)^2F_N(t_2)^2,
\label{cross2}
\ee
where
\ba
t_1&=&(p_1-p_1^\prime)^2\nn\\
t_2&=&(p_2-p_2^\prime)^2,
\nn
\ea         
and $F_N(t)$ is the strong form factor of the nucleon for which we
will use
the electromagnetic form factor \ct{dl}
\be
 F_N(t)=\frac{4m_p^2-2.79t}{(4m_p^2-t)(1-t/0.71)^2}.
\la{ff1}
\ee

 In the numerical calculation we use the NLO approximation for the strong
 coupling constant
\begin{equation}
\alpha_s(\rho)=-\frac{2\pi}{\beta_1t}(1+\frac{2\beta_2log t}{\beta_1t}) ,
\label{e17}
\end{equation}
 where
\begin{equation}
\beta_1=-\frac{33-2N_f}{6},\ \beta_2=-\frac{153-19N_f} {12}\nonumber
\end{equation}
 and
\begin{equation}
t=log(\frac{1}{\rho_0^2\Lambda^2}+\delta).
\label{e18}
\end{equation}
 In Equation (\ref{e18}), the parameter,
$\delta\approx1/\rho_0^2\Lambda^2$,
 provides a smooth interpolation of the value of  $\alpha_s(\rho)$
 from the perturbative $(\rho\rightarrow0)$ to the nonperturbative
 regime $(\rho\rightarrow\infty)$~\cite{a8},
and  for $N_f=3$ we have used $\Lambda=230\; MeV$, and $\rho_0=1.6 \; GeV^{-1}$.

The exact phase space integration of  \re{cross} gives
the total cross section of the reaction \re{reaction} as  
a function of the excess energy in c.m. system 
$Q=\sqrt{S}-m_n-m_p-m_\eta$, 
$S=(p_1+p_2)^2$, and
is presented in Fig.3 for the different 
values of the instanton size $\rho_c$.  
This dependence is in qualitative agreement with
experimental data of WASA/PROMICE Collaboration at CELSIUS 
\ct{cel}. It should be mentioned that the instanton 
induced cross section of the $\eta$ production in $pn$ scattering 
is very sensitive to the average size of instanton in the QCD vacuum 
and for $\rho_c=2.1 \; GeV^{-1}$ it describes the experimental data.
\begin{figure}[htb]
\centering
\vskip -.5cm

\centering
\epsfig{file=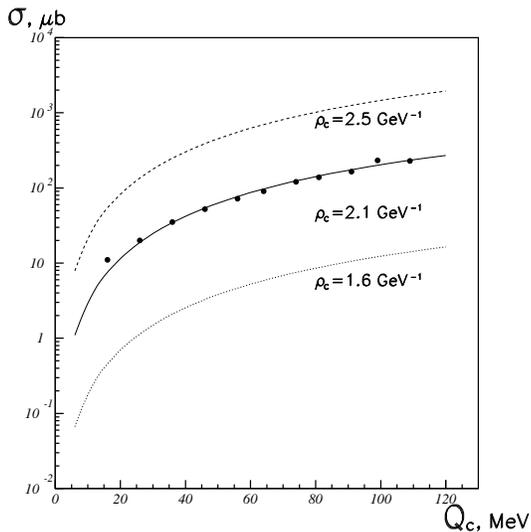,width=7.5cm}
\caption{\it $Q$-dependence of 
 $\eta$   production cross section for the different values of the 
average instanton size in comparison with
experimental data of WASA/PROMICE Collaboration.}
\end{figure}

Before ending this section it is worth stressing the main difference
between our approach and the 
hadronic approach. In the latter  quantum numbers and energy
considerations determine the 
intermediate meson, the $S_{11}$, whose coupling constant is at
present unknown and whose properties, 
mass and width, have to be feed in from experiments. In our approach
the calculation is basically 
parameter free, despite the fact that we have allowed some freedom
into the size parameter
reflecting the crudeness of the approximations involved. An
independent determination of the 
$S_{11}$ coupling constant would clarify the relation between the
nonperturbative microscoscopic 
mechanism presented here and the hadronic approach. The conclusions
drawn thus far 
are only slightly affected by  final state interactions \ct{fsi} due
to their  structure and  
magnitude compared with that of the  of the enhancement.

\section{Conclusion}

We have shown that a novel mechanism for $\eta$ production in 
nucleon-nucleon scattering might be behind the anomalous enhancement
found for proton-neutron collisions. The proposed mechanism  
is related with the  presence of strong nonperturbative 
fluctuations of gluons fields in the QCD vacuum, called instantons.
It has been shown that the instanton induced quark-gluon interaction,
due specially to its singular flavor dependence, 
is able to explain the  large $\eta$-meson production cross section   
near threshold in proton-neutron scattering. The calculation
has been realized within a simplified scenario of
baryon structure, the diquark description, which maximizes the flavor
dependence of the interaction. However it is immediate that other
models will not avoid the basic mechanism, whose origin is solely in
the structure of the instanton interaction.

The final result for the cross section of the $\eta$ production is 
very sensitive to the value of the instanton size. This
observation opens the door to investigate the parameters of the QCD 
vacuum in the threshold production of $\eta$ (and other mesons) 
with high accuracy. 

Naturally our proposed mechanism appears in nature entangled with other
ones. Their experimental separation is very important.
A possible way is the determination  of the energy and angular
dependence of the production processes. Due to the point-like origin 
of the instanton-induced interaction, its dependences are totally
different from those of other mechanisms. For example in 
$\eta$-meson production the competing mechanism is the production of
an intermediate $S_{11}$-resonance and certainly its energy and
angular dependence are very different. A very clean way to disentangle 
the various mechanisms is the measurement of the {\it spin} 
dependence of the cross section in the production processes by using 
polarized nucleon beams and targets. In this case the instantons give 
a large contribution to the double spin asymmetries due to 
very strong helicity dependence of t'Hooft's interaction 
\re{et1}.

\section*{Acknowledgements}
One of us (V.V.) thanks L. Alvarez-Ruso for tuition in the subject of
final state interactions and for useful comments. This work was partially 
supported by DGICYT grant \# PB97-1227.

\end{document}